\documentstyle{article}
\font\cero=cmss10 scaled 1728
\font\uno=cmssbx10 scaled 1200
\begin{document}
\begin{flushleft}
{\cero Local continuity laws on the phase space of Einstein
equations with sources} \\[3em]
\end{flushleft}
{\sf R. Cartas-Fuentevilla}\\
{\it Instituto de F\'{\i}sica, Universidad Aut\'onoma de Puebla, Apartado
Postal J-48 7250 Puebla, Pue., M\'exico, and
Enrico Fermi Institute, University of Chicago, 5640 S. Ellis Ave.,
Chicago, Illinois 60637}

Local continuity equations involving background fields and variations of
the fields, are obtained for a restricted class of solutions of the
Einstein-Maxwell and Einstein-Weyl theories using a new approach based on
the concept of the adjoint of a differential operator. Such covariant
conservation laws are generated by means of decoupled equations and their
adjoints in such a way that the corresponding covariantly conserved
currents possess some gauge invariant properties and are expressed in
terms of Debye potentials. These continuity laws lead to both a covariant
description of bilinear forms on the phase space and the existence of
conserved quantities. Differences and similarities with other approaches,
and extensions of our results are discussed.

\noindent Running title: Local continuity laws....

\begin{center}
{\uno I. INTRODUCTION}
\end{center}

The study of the conservation laws is present in practically all areas of
physics. In the context of the theories involving gravity, such a study
becomes particularly interesting because of the lack of conserved currents
representing the conservation of energy and momentum. Moreover, in the
covariant description of a symplectic structure on the phase space of
classical systems, a
covariantly conserved two-form on such phase space is required. Such
descriptions yield a canonical formalism manifestly covariant in field
theory, which represents a starting point for a quantum mechanical
treatment \cite{1}. More specifically, in this context, the phase space
is defined as the manifold of solutions of the classical equations of
motion. Such a definition has the advantage of manifest covariance
\cite{1}. The existence of a two-form (say $J^{\mu}$), covariantly
conserved ($\partial_{\mu} J^{\mu} = 0$) on such phase space, lead to a
symplectic structure $\omega$ defined as $\omega \equiv \int_{\Sigma}
J^{\mu} d\Sigma_{\mu}$ (where $\Sigma$ is an initial value hypersurface),
independent of the choice of $\Sigma$. Moreover, the condition that
Poisson brackets satisfy the Jacoby identity, holds if $\omega$ is closed
(which means that $\omega$ is a two-form whose exterior derivative
vanishes). This condition is satisfied if $J^{\mu}$ itself is closed.
Several methods have been proposed in the literature for constructing
these conserved currents, from the guess for the explicit form of the
currents \cite{1},  to the very general result on the existence of a
conserved current (the symplectic current) for the field variations for
any field theory derived from a Lagrangian action \cite{1,2}.
Alternatively, equivalent conserved currents have been obtained directly
from the equations for the field variations using the fact that there
exists a conserved current associated with any system of homogeneous
linear partial differential equations that can be written in terms of a
self-adjoint operator \cite{3,4}. Such results have been obtained without
any statement about the background fields. Nevertheless, the currents
constructed in these ways are associated with the variations of the gauge
fields. For example, in the case of the Einstein-Maxwell theory, such
currents are expressed in terms of the metric and vector potential
variations, and in this manner are defined modulo gauge transformations
[1-4]. However, as it will be shown in the present paper, one can construct,
for a restricted class of spacetimes, conserved currents (and the
geometrical structures associated) without the gauge freedom of the matter
fields. The approach is based on the concept of the adjoint of a
differential operator, which represents an extension of that employed in
Refs. \cite{3,4}. Our starting point is the existence of appropriate
decoupled equations for the restricted class of spacetimes under study,
which have been obtained using the Newman-Penrose formulation \cite{5,6}. It
is shown that the conserved currents are also independent of the perturbed
tetrad gauge freedom, which appears when the variations of the
Newman-Penrose quantities are involved.

A previous application of the present approach has been given in the study
of the perturbations of the Schwarzschild black hole, leading up to an
important result on the existence of conservations laws for non-Hermitian
systems, which has a direct extension to the scheme of string theory
\cite{7}. In this manner, another of the purposes of this article is to
establish conservation laws for more general spacetimes, analogous to that
appearing in Ref.\ \cite{7}. However, unlike the results given in that
reference, one of the main novelties in the present article is the
interpretation of the corresponding conserved currents as bilinear forms
on the phase space, which is connected with the very important concept of
symplectic structures. This subject will be extended later on.

The outline of this paper is as follows. Section\ II is dedicated to
establish the general relationship between adjoint operators and conserved
currents. The application of this very general result for identifying the
conserved currents for the field variations in the Einstein-Maxwell and
Einstein-Weyl theories is given in Sec.\ III. The properties of such
conserved currents, and their relations with the existence of symplectic
structures on the phase space are discussed. We finish in Sec.\ IV with
some concluding remarks and future extensions of the present results. \\[2em]

\noindent {\uno II. ADJOINT OPERATORS AND CONSERVED CURRENTS}
\vspace{1em}

{\bf New branch}: In Refs.\ \cite{3,4} it has been shown that there exists
a conserved current associated with any system of homogeneous linear
partial differential equations that can be written in terms of a
self-adjoint operator. This result is limited for a self-adjoint system,
for which the corresponding conserved current depends on a pair of
solutions admitted by such a system. However, as we shall see below, there
exists a more general possibility that extends for systems of equations
which are not self-adjoint necessarily. The demonstration is very easy,
following the basic idea of Refs. \cite{3,4}:

In accordance with Wald's definition \cite{7}, if ${\cal E}$ corresponds
to a linear partial differential operator which maps $m$-index tensor fields
into $n$-index tensor fields, then, the adjoint operator of $\cal E$,
denoted by ${\cal E}^{\dag}$, is that linear partial differential operator
mapping $n$-index tensor fields into $m$-index tensor fields such that
\begin{equation}
     g^{\rho\sigma\ldots} [{\cal E}(f_{\mu\nu\ldots})]_{\rho\sigma\ldots} -
     [{\cal E}^{\dag}(g^{\rho\sigma\ldots})]^{\mu\nu\ldots} f_{\mu\nu\ldots}
     = \nabla_{\mu}  J^{\mu},
\end{equation}
where $J^{\mu}$ is some vector field depending on the fields $f$ and $g$.
From Eq.\ (1) we can see that this definition automatically guarantees
that, if the field $f$ is a solution of the linear system ${\cal E} (f) =
0$ and $g$ a solution of the adjoint system ${\cal E}^{\dag} (g) = 0$,
then we obtain the local continuity law $\nabla_{\mu}J^{\mu}=0$, which
establishes that $J^{\mu}$ is a covariantly conserved current. This fact
means that for any homogeneous equation system, one can always construct a
conserved current taking into account the adjoint system. This general
result contains the self-adjoint case $({\cal E}={\cal E}^{\dag})$ as a
particular one, for which $f$ and $g$ correspond to two independent
solutions.

In the present work, $f$ and $g$ will be associated  with the first-order
variations of the backgrounds fields. Such field variations will
correspond, on the phase space, to one-forms [1]. In this manner, the
left-hand side of Eq.\ (1) can be understood as a wedge product on such
phase space: $g\wedge {\cal E}(f) - {\cal E}^{\dag}(g)\wedge f=
\nabla_{\mu}J^{\mu}$, and something similar for the bilinear form
$J^{\mu}$ in its dependence on the fields $f$ and $g$ (the operators
${\cal E}$, ${\cal E}^{\dag}$, and $\nabla_{\mu}$ are depending only on
the background fields).

{\bf Traditional branch}: It is important, for completeness, that we
outline the original idea for introducing the definition (1) in Ref.\
\cite{7}: reduction of systems of linear partial differential equations to
equations for scalar potentials (called Debye potentials), which determine
a complete solution of the original system.

If we have the linear system ${\cal E}(f) = 0$, and there exist linear
operators such that
\[
   {\cal S} {\cal E} = {\cal O} {\cal T}, \nonumber
\]
identically, then the field ${\cal S}^{\dag} (\psi)$ satisfies the
equation
\[
   {\cal E}^{\dag} ({\cal S}^{\dag}(\psi)) = 0, \nonumber
\]
provided that the scalar field $\psi$ satisfies
\[
   {\cal O}^{\dag} (\psi) = 0. \nonumber
\]

In particular, if ${\cal E}$ is self-adjoint $({\cal E}^{\dag} = {\cal
E})$, then $f = {\cal S}^{\dag} (\psi)$ is a solution of ${\cal E} (f) =
0$. For example, in the case considered in the present work, the
(matrix) operator governing the field perturbations in the
Einstein-Maxwell and Einstein-Weyl theories are, in fact, self-adjoint
\cite{3,4}.

Moreover, the existence of operators ${\cal S}$, ${\cal O}$, and ${\cal
T}$ satisfying the above identity, is equivalent to the existence of a
decoupled system
\[
   {\cal O} (\Psi) = 0, \nonumber
\]
obtained from the original system ${\cal E} (f) = 0$, such that the scalar
field $\Psi = {\cal T} (f)$.

Now, we can mix both branches of the adjoint operators scheme: since the
fields $\psi$ and $\Psi$ satisfy equation which are adjoints to each
other, we can
establish, in according to the first branch, that
\[
   \psi ({\cal O} \Psi) - ({\cal O}^{\dag} \psi) \Psi = \nabla_{\mu}
{\cal J}^{\mu} (\psi, \Psi), \nonumber
\]
which means that $\nabla_{\mu} {\cal J}^{\mu} (\psi, \Psi) = 0$.
Furthermore, since $\Psi$ is finally depending on $\psi$ $\left( \Psi =
{\cal T} (f) = {\cal T} ({\cal S}^{\dag} (\psi)) \right)$, ${\cal
J}^{\mu}$ is dependent only on $\psi$ (however, see the paragraph before
Eq.\ (8)).

On the other hand, although this result has been established assuming only
tensor fields and the presence of a single equation, such a result can be
extended in a direct way to equations involving spinor fields, matrix
fields, and the presence of more than one field \cite{8}.

It is worth pointing out some issues on the notation. The first-order
field variations appearing in Ref.\ [5,6] are denoted by a superscript
B. On the other hand, the field variations coincide, in according to
Witten's interpretation \cite{1}, with an infinite-dimensional
generalization of the usual exterior derivative, which is traditionally
represented by the symbol $\delta$. However, in Ref.\ [5,6], the
Newman-Penrose formalism is used, in which the symbol $\delta$ is employed
for denoting one of the directional derivatives defined by the null
tetrad.
In this manner, for avoiding confusion, we will maintain the symbol
$\delta$ as usual in the Newman-Penrose notation, and the superscript B
for the first-order field variations (the exterior derivative of
background fields). In the present article, the exterior derivative will
not be performed explicity, and it will be sufficient for our purposes to
understand any quantity with the superscript B as a one-form on the phase
space. Quantities without such a superscript will correspond to
background fields, which mean zero-forms on the phase space \cite{1}. With
these previous considerations, formulae and notation of Ref.\ [5,6] will
be used throughout this paper.  \\[2em]

\noindent {\uno III. COVARIANT CONTINUITY LAWS}
\vspace{1em}

In Ref. \cite{5}, the metric and the vector potential
variations of the Einstein-Maxwell equations have been obtained starting
from a decoupled set of equations governing certain combinations of field
variations, and have the form:
\begin{equation}
     {\cal O}(\Psi^{\rm B}) = 0,
\end{equation}
where the $4\times4$ matrix operator ${\cal O}$ can be read from Eq.\ (24)
of that reference, and $(\Psi^{\rm B})$ is the matrix:
\begin{equation}
     (\Psi^{\rm B}) = \pmatrix{\Psi_{0}^{\rm B} \cr - \tilde{\Psi}_{1}^{\rm
     B} \cr \phi_{1} \tilde{\kappa}^{\rm B} \cr - \phi_{1}
     \tilde{\sigma}^{\rm B}},
\end{equation}
where
\begin{eqnarray}
     \tilde{\Psi}^{\rm B}_{1} \!\! & \equiv \!\! & (2 \phi_{1} \Psi^{\rm
     B}_{1} - 3 \Psi_{2} \phi^{\rm B}_{0})/ 2 \phi_{1}, \nonumber \\
     \tilde{\kappa}^{\rm B} \!\! & \equiv \!\! & \kappa^{\rm B} + (D - 2
     \epsilon - \rho) (\phi^{\rm B}_{0}/ 2 \phi_{1}), \\
     \tilde{\sigma}^{\rm B} \!\! & \equiv \!\! & \sigma^{\rm B} + (\delta
     - 2 \beta - \tau) (\phi^{\rm B}_{0}/ 2 \phi_{1}). \nonumber
\end{eqnarray}
The decoupled system (2) was obtained provided that the background
electromagnetic field is non-null with one of its principal null directions
being geodetic and shear free, which means that $\Psi_{0} = 0 = \Psi_{1}$,
$\kappa = 0 = \sigma$, and $\phi_{0} = 0$ in the background \cite{5}. Note
that the variations of these vanishing background quantities are those
involved in $(\Psi^{\rm B})$. As described in previous Section, the
variations of background fields correspond to one-forms, in particular,
those ones in Eq.\ (4). However, as shown in Appendix B, they correspond
to exact one-forms. This property will be useful below.

On the other hand, according to Wald's method used in Ref. \cite{5}, the
matrix potential $(\psi)$, which determines the complete variations,
satisfies the equations (see Eqs.\ (25) and (27) of Ref. \cite{5})
\begin{equation}
     {\cal O}^{\dag} (\psi) = 0,
\end{equation}
where
\begin{equation}
     (\psi) = \pmatrix{M_{1'} \cr - M_{0'} \cr \psi_{\rm G} \cr - \psi_{\rm
     E}}.
\end{equation}
Thus, from the definition (1) we have that $(\psi)\wedge{\cal O}
({\Psi}^{\rm B}) - {\cal O}^{\dag}(\psi)\wedge({\Psi}^{\rm B}) =
\nabla_{\mu} J^{\mu} ({\Psi}^{\rm B}, {\psi})$. The left-hand side contains
terms of the form $M_{1'}\wedge{\cal O}_{11}(\Psi_{0}^{\rm B}) - {\cal
O}_{11}^{\dag}(M_{1'})\wedge\Psi_{0}^{\rm B}$, where ${\cal O}_{11} \equiv
\overline{\delta}-4\alpha+\pi$ and ${\cal O}_{11}^{\dag}\equiv
(\overline{\delta}-4\alpha+\pi)^{\dag}=\overline{\delta} + 3\alpha
+\overline{\beta}-\overline{\tau}$, are the entries 11 of the
$4\times 4$ matrices ${\cal O}$ and ${\cal O}^{\dag}$ respectively (see
Eqs.\ (24) and (25) in Ref.\ \cite{5}). Taking into account that
$\overline{\delta}\equiv\overline{m}^{\mu}\partial_{\mu}$, and that is
acting on scalar fields, such a term can be expressed in the form
$M_{1'}\wedge{\cal O}_{11}(\Psi_{0}^{\rm B}) - {\cal O}_{11}^{\dag}(M_{1'})
\wedge\Psi_{0}^{\rm B}=\nabla_{\mu} (\overline{m}^{\mu}M_{1'}\wedge
\Psi^{\rm B}_{0})$, and similarly for the remaining terms considering that
the tetrad vectors $D$, $\Delta$, and $\delta$ are defined by $l^{\mu}
\partial_{\mu}$, $n^{\mu}\partial_{\mu}$, and $m^{\mu}\partial_{\mu}$
respectively. In this manner, from Eqs.\ (2) and (5) we obtain explicity
the following covariant conservation law:
\begin{eqnarray}
     \nabla_{\mu}J^{\mu}=0, \qquad J^{\mu} \equiv - l^{\mu} (M_{1'}
\wedge\tilde{\Psi}^{\rm B}_{1} + 2 \phi_{1} \psi_{\rm G}\wedge
\tilde{\sigma}^{\rm B}) - n^{\mu} (M_{0'} \wedge \Psi^{\rm B}_{0}
+ 2 \phi_{1} \psi_{\rm E}\wedge \tilde{\kappa}^{\rm B}) \nonumber\\
     + m^{\mu} (M_{0'}\wedge \tilde{\Psi}^{\rm B}_{1} + 2 \phi_{1}
\psi_{\rm G} \wedge \tilde{\kappa}^{\rm B}) + \overline{m}^{\mu} (M_{1'}
\wedge \Psi^{\rm B}_{0} + \phi_{1} \psi_{\rm E}\wedge \tilde{\sigma}^{\rm B}).
\end{eqnarray}

It is very easy to determine that, like $(\Psi^{\rm B})$ in Eq.\ (3),
$(\psi)$ in Eq.\ (6) is made out of one-forms. From Eqs. (28) and (29)
of Ref.\ \cite{5}, the metric and vector potential variations are given by
\[
     \pmatrix{(h_{\mu\nu}) \cr (b_{\mu})} \equiv \pmatrix{(g_{\mu\nu})^{\rm B}
     \cr (A_{\mu})^{\rm B}} = {\cal S}^{\dag} (\psi), \nonumber
\]
where ${\cal S}^{\dag}$ is an operator depending only on background fields
(see Eqs.\ (26), (28), and (29) in Ref.\ \cite{5}), which correspond to
zero-forms. Since $h_{\mu\nu}$ and $b_{\mu}$ correspond to one-forms, from
the preceding equation, $(\psi)$ correspond also to one-forms. This
implies automatically that $J^{\mu}$ in Eq.\ (7) is a (non-degenerate)
two-form on the corresponding phase space. However, we have to demonstrate
that $J^{\mu}$ corresponds to a closed two-form for obtaining a
symplectic structure on the phase space. For this purpose, we use the
results in Appendix B. For example, the first term $-l^{\mu} M_{1'} \wedge
\tilde{\Psi}^{\rm B}_{1}$ for $J^{\mu}$ in Eq.\ (7) can be rewritten as:
\[
    -l^{\mu} M_{1'} \wedge \tilde{\Psi}^{\rm B}_{1} = \left[ l^{\mu}
    M_{1'} \left( \Psi_{1} - \frac{3\Psi_{2}}{2\phi_{1}} \phi_{0} \right)
    \right]^{\rm B}, \nonumber
\]
where Eq.\ (B1) has been used; the above expression
means that such a term is an exact two-form (and automatically a
closed two-form). Similarly, using Eqs.\ (B1) and (B2), the remaining
terms in Eq.\ (7) correspond to closed two-forms. Therefore, $J^{\mu}$
itself is a closed two-form, as required for constructing a symplectic
structure. Such geometrical structure is defined by $\omega \equiv
\int_{\Sigma} J^{\mu}d\Sigma_{\mu}$, where $\Sigma$ is an initial value
hypersurface. As $J^{\mu}$ is conserved, $\omega$ is independent of the
choice of $\Sigma$ and, in particular, is Poincar\'e invariant \cite{1}.
In this manner, $\omega$ is a covariant symplectic structure on the phase
space.

On the other hand, as shown in the Appendix A, the one-forms
$\Psi_{0}^{\rm B}$, $\tilde{\Psi}_{1}^{\rm B}$, $\tilde{\kappa}^{\rm B}$,
and $\tilde{\sigma}^{\rm B}$, are invariant under gauge transformations of
the vector potential variations, $b_{\mu} \rightarrow b_{\mu} + \nabla_{\mu}
\chi$, and thus, $J^{\mu}$ and $\omega$ have the same invariance property
(in contrast with those of Refs.\ [1-4]). In field space, $\nabla_{\mu}
\chi$ corresponds to the gauge directions, therefore, this invariance
property defines automatically a gauge-invariant two-form $\omega$ on the
subtler space $Z\equiv \hat{Z}/G$, being $\hat{Z}$ the space of solutions
and $G$ the group of gauge transformations \cite{1}. Similarly, since such
one-forms are also invariant under infinitesimal rotations of the tetrad
(see Appendix A), we have also a two-form $\omega$ independent on the
perturbed tetrad gauge freedom, defined on the subtler space $T \equiv
\hat{Z}/H$, being $H$ the group of infinitesimal rotations of the
perturbed tetrad. However, $J^{\mu}$ and $\omega$ are not gauge invariant
with respect to infinitesimal diffeomorphisms (in similarity to those of
Refs.\ [1-4]).

From Eq.\ (7) we can see that the conserved current depends on the
background fields and the solutions admitted by the decoupled system and its
adjoint. However, the entries of the matrix $(\Psi^{\rm B})$ can be defined
in terms of the metric and vector potential variations, which in turn, are
defined in terms of the potential matrix $(\psi)$ (see Eqs.\ (28) and (29)
of Ref. \cite{5}), and therefore, $J^{\mu}$ and $\omega$ can be expressed
in terms of a single complex solution of the equations for the potential
$(\psi)$. However, in the more general case, if $(\psi)_{1}$ is a solution
admitted by the equations for the potentials, the matrix $(\Psi^{\rm B})$
can be expressed in terms of a second solution $(\psi)_{2}$, in general
different of $(\psi)_{1}$. Therefore, $J^{\mu}$ and $\omega$ are expressed
in terms of a pair of solutions for those equations. Finally, $J^{\mu}$
and $\omega$ will be given completely in terms of the potentials
(one-forms), which become the fundamental geometrical structures. In this
manner, the analysis of the structure of the phase space has been reduced
to the study of equations for the potentials, which is a relatively simple
issue.

Similarly, when the background electromagnetic field is null (and there is a
possibly nonzero cosmological constant) the continuity equation for the field
variations in the Einstein-Maxwell theory is given by (see Eqs.\ (21) and
(24) of Ref.\ \cite{6})
\begin{eqnarray}
    \nabla_{\mu}J^{\mu}=0, \qquad
J^{\mu}\equiv l^{\mu}[\psi_{\rm G}\wedge (\Delta - 4 \gamma + \mu)
     \Psi^{\rm B}_{0} + \psi_{\rm E}\wedge(\Delta - 2 \gamma + \mu)
     \varphi^{\rm B}_{0} \nonumber \\
+ 2 \overline{\varphi}_{2} (\psi_{\rm G}\wedge(D - 2\epsilon)
     \varphi^{\rm B}_{0} -(D + 4 \epsilon + 3 \rho) \psi_{\rm G}\wedge
     \varphi^{\rm B}_{0} )] \big. \nonumber \\
- n^{\mu} [(D + 4 \epsilon + 3 \rho) \psi_{\rm G} \wedge\Psi^{\rm B}_{0}
     + (D + 2 \epsilon + \rho) \psi_{\rm E} \wedge\varphi^{\rm B}_{0}]
     \nonumber \\
- m^{\mu} [\psi_{\rm G}\wedge (\overline{\delta} - 4 \alpha + \pi) \Psi^{\rm
     B}_{0} + \psi_{\rm E}\wedge (\overline{\delta} - 2 \alpha + \pi)
     \varphi^{\rm B}_{0} ] \nonumber \\
+ \overline{m}^{\mu} [ (\delta + 4 \beta + 3 \tau) \psi_{\rm G}\wedge
     \Psi^{\rm B}_{0} + (\delta + 2 \beta + \tau) \psi_{\rm E}\wedge
     \varphi^{\rm B}_{0} ],
\end{eqnarray}
where similarly $J^{\mu}$ is a closed two-form on the phase space, and the
corresponding symplectic structure $\omega \equiv \int_{\Sigma} J^{\mu}
d\Sigma_{\mu}$ is Poncair\'e invariant because $J^{\mu}$ is conserved.
Furthermore, $J^{\mu}$ and the corresponding symplectic form $\omega$ have
essentially
the same invariance properties that the previous case (see Appendix A),
and  therefore are defined in a gauge-invariant form on the corresponding
subtler spaces $Z$ and $T$. Analogously, they are given, in general, in
terms of a pair of solutions admitted by the corresponding equations for
the potentials (see Eqs.\ (24) in Ref.\ \cite{6}).

In the case of solutions of the Einstein-Weyl equations such that the flux
vector of the neutrino field is tangent to a shear-free congruence of null
geodesics, the field variations satisfy the continuity equation (see
Eqs.\ (38)-(47) of Ref. \cite{5}):
\begin{eqnarray}
      \nabla_{\mu} \big\{ l^{\mu}[-M_{1'}\wedge \tilde{\Psi}^{\rm
B}_{1} - \eta_{1} \psi_{\rm G}\wedge\tilde{\sigma}^{\rm B} - i k \eta_{1}
\overline{\eta}_{1'} M_{1'}\wedge \tilde{\kappa}^{\rm B} + 2 i k \eta_{1}
\overline{\eta}_{1'} M_{0'}\wedge \tilde{\sigma}^{\rm B}] \nonumber\\
      - n^{\mu} [ M_{0'}\wedge \Psi^{\rm B}_{0} + \eta_{1} \psi_{\rm
N}\wedge \tilde{\kappa}^{\rm B} ] \big. \nonumber \\
     + m^{\mu} [\eta_{1} \psi_{\rm G}\wedge \tilde{\kappa}^{\rm B} +
M_{0'}\wedge \tilde{\Psi}^{\rm B}_{1} - i k \eta_{1} \overline{\eta}_{1'}
M_{0'} \wedge \tilde{\kappa}^{\rm B}] + \overline{m}^{\mu} [ M_{1'}\wedge
\Psi^{\rm B}_{0} + \eta_{1} \psi_{\rm N}\wedge \tilde{\sigma}^{\rm B} ] \big.
\big\} = 0.
\end{eqnarray}
where $M_{0'}$, $M_{1'}$, $\psi_{N}$, and $\psi_{G}$ are the potentials,
and similarly $\Psi^{\rm B}_{0}$, $\tilde{\Psi}^{\rm B}_{1}$,
$\tilde{\sigma}^{\rm B}$, and $\tilde{\kappa}^{\rm B}$ are defined in
terms of such potentials (see Eqs.\ (47) of Ref.\ \cite{5}).

The continuity equations (7)--(9) deserve some additional comments. The
existence of such continuity laws follows directly from the definition of
the adjoint of a differential operator, and we have only to identify
conveniently the system of differential equations for applying such
definition. In this sense, the adjoint operators scheme allows us {\it to
reveal} the underlying geometrical structures for the corresponding
theory, in a elementary and straightforward way. The argument involved
is, as seen above, very simple.

On the other hand, in the Lagrangian approach known in the literature
\cite{2}, the symplectic currents arise from the exterior derivative of
the so called {\it symplectic potentials}, which correspond to
one-forms on the phase space. This guarantees that such bilinear forms
correspond automatically to  closed two-forms (symplectic structures) on
the phase space. However, in the present approach, the symplectic currents
in Eqs.\ (7)--(9) are constructed in a direct way using only the concept of
adjoint operators, regardless of the existence of symplectic potentials.
Furthermore, our symplectic currents are generated from the usual
derivatives (essentially directional derivatives) of the potentials, which
correspond also to one-forms on the phase space. This suggests a possible
relationship between the fundamental structures of the different
approaches: the symplectic potentials of the Lagrangian approach and the
potentials considered in the present scheme. This may be the subject of
forthcoming works.
 \\[2em]

\begin{center}
{\uno V. CONCLUDING REMARKS}
\end{center}
\vspace{1em}
As we have seen, the present approach based on the concept of the adjoint
of a differential operator, represents a more direct and elementary
procedure for constructing geometrical structures manifestly covariant on
the phase space. Specifically, it  has allowed us to demonstrate that,
for the restricted class of spacetimes considered here, there exist two
covariantly conserved currents for the field variations of each theory,
the symplectic current (which coincides fully with that obtained from the
Lagrangian approach \cite{3,4}), and that found in the present approach
(which is no obtained directly from a Lagrangian approach). Since both
currents can be expressed, finally, in terms of solutions of the equations
for the corresponding scalar potentials, our fundamental geometrical
structures, a direct comparison is possible and convenient in order to
establish the differences and similarities. On the other hand, although
some restrictions on the background fields have been required, a lot of
solutions of wide interest is contained in the cases considered here:
rotating (charged) black holes, colliding plane waves, etc. The detailed
study of the corresponding currents and geometrical structures associated
with those particular solutions will be the subject of subsequent works. In
fact, as mentioned in the Introduction, works along these lines have
permitted to demonstrate that, the continuity equation (8) directly yields,
in the particular case of the gravitational and electromagnetic
perturbations of the Schwarzschild black hole, the conservation of the
energy for those perturbations without invoking, as usually, the
Schr\"odinger-type equation for establishing such a conservation
relation. This result is connected with the existence of conserved
quantities for non-Hermitian system of differential equations \cite{7}.

Finally, the general result established in Sec.\ II can be understood as an
important extension of the original Wald's method \cite{9}: wherever there
exists an appropriate decoupled equation, it is not only possible to express
the complete solution in terms of scalar potentials, but also to find
automatically a corresponding (covariantly) conserved current, which can
be expressed, in turn, in terms of the same potentials. The possible
applications of such a result in the modern theories involving gravity and
matter fields such as string theory (and other areas of physics) are open
questions \cite{7}. \\[2em]

\begin{center}
{\uno ACKNOWLEDGMENTS}
\end{center}
\vspace{1em}

This work was supported by CONACyT and the Sistema Nacional de
Investigadores
(M\'exico). The author wants to thank Dr. G. F. Torres del Castillo for his
comments on the manuscript and suggestions, and Professor Robert Wald for
the kind hospitality provided at the Enrico Fermi Institute, University of
Chicago. \\[2em]

 \begin{center}
{\uno APPENDIX A: INVARIANCE PROPERTIES}
\end{center}
\renewcommand{\theequation}{A\arabic{equation}}
\setcounter{equation}{0}
\vspace{1em}

Although it is not mentioned explicitly in Refs. \cite{5,6}, the perturbed
quantities appearing in the decoupled equations are invariant under
infinitesimal rotations of the tetrad. For example, in the case of the
perturbed quantity $\tilde{\kappa}^{\rm B}$ (see Eqs.\ (4)) made out of the
perturbations of the spin coefficient $\kappa$ and of the component of
electromagnetic field $\phi_{0}$, we have that:
\begin{eqnarray}
     \kappa^{\rm B} \!\! & \equiv & \!\!  -(l^{\mu} l^{\nu} \nabla_{\nu}
     m_{\mu})^{\rm B} = l^{\mu} l^{\nu} m_{\gamma}
     (\Gamma^{\gamma}_{\mu\nu})^{\rm B}-l^{\mu}l^{\nu} \nabla_{\nu}
     m_{\mu}^{\rm B} - (l^{\mu} l^{\nu})^{\rm B} \nabla_{\nu} m_{\mu}
     \nonumber \\
     & \!\! = \!\! & l^{\mu}l^{\nu}m_{\gamma}(\Gamma^{\gamma}_{\mu\nu})^{\rm
     B} - \frac{1}{2} (\overline{\pi} - \tau) l^{\mu} l^{\nu} h_{\mu\nu} -
     (D - 2 \epsilon - \rho) (l^{\mu} m^{\rm B}_{\mu}),
\end{eqnarray}
where $(\Gamma^{\gamma}_{\mu\nu})^{\rm B} = \frac{1}{2} g^{\gamma\rho}
(\nabla_{\mu} h_{\nu\rho} + \nabla_{\nu} h_{\mu\rho} - \nabla_{\rho}
h_{\mu\nu})$ and $h_{\mu\nu}$ is the metric perturbation. In this manner,
the first and second terms in the above equation are defined completely in
terms of $h_{\mu\nu}$. On the other hand, $l^{\mu} m^{\rm B}_{\mu}$ is
dependent on the perturbed tetrad gauge freedom. Furthermore, from the
definition $\phi_{0}\equiv l^{\mu}m^{\nu}F_{\mu\nu}$, one finds that
\begin{eqnarray}
     \phi_{0}^{\rm B} = l^{\mu} m^{\nu} F^{\rm B}_{\mu\nu} - (\phi_{1} +
     \overline{\phi_{1}}) l^{\mu} m^{\nu} h_{\mu\nu} + 2\phi_{1} (l^{\mu}
     m^{\rm B}_{\mu}),
\end{eqnarray}
where $\phi_{1}$ is the only nonvanishing component of the background
electromagnetic field, and $F^{\rm B}_{\mu\nu} = \partial_{\mu} b_{\nu} -
\partial_{\nu} b_{\mu}$, being $b_{\mu}$ the vector potential perturbations.
Therefore, from Eqs.\ (A1) and (A2) we can see easily that the quantity
$\tilde{\kappa}^{\rm B} = \kappa^{\rm B} + (D - 2 \epsilon - \rho)
(\phi^{\rm B}_{0}/ 2 \phi_{1})$ is independent of the perturbed tetrad
gauge freedom. Similarly for the remainding quantities $\Psi^{\rm B}_{0},
\tilde{\Psi}^{\rm B}_{1}$, and $\tilde{\sigma}^{\rm B}$ in Eq.\ (4). Thus,
$(\Psi^{\rm B})$ is invariant under the transformation considered. Since the
field perturbation $F_{\mu\nu}^{\rm B}$ is invariant under the ordinary
gauge transformation $b_{\mu} \rightarrow b_{\mu} + \nabla_{\mu }\chi$,
where $\chi$ is an arbitrary scalar field, $\phi_{0}^{\rm B}$ in Eq.\ (A2)
is also invariant under this transformation and, in this manner, ($\Psi^{\rm
B}$) is (see Eqs.\ (4)). However, it is not difficult to demonstrate that
($\Psi^{\rm B}$) is not invariant under infinitesimal diffeomorphisms
$b_{\mu} \rightarrow b_{\mu} + {\cal L}_{\chi} A_{\mu}$, and $h_{\mu\nu}
\rightarrow h_{\mu\nu} + {\cal L}_{\chi} g_{\mu\nu}$, where $\chi^{\mu}$ is
an arbitrary vector field, and $A_{\mu}$ and $g_{\mu\nu}$ the background
fields.

Similarly, in the case of solutions of the Einstein-Maxwell theory with a
null background electromagnetic field ($\phi_{2}$ is the only nonvanishing
component) \cite{6}, we find that
\begin{eqnarray}
     \phi_{0}^{\rm B} = l^{\mu} m^{\nu} F^{\rm B}_{\mu\nu} + \frac{1}{2}
     \overline{\phi_{2}} l^{\mu} m^{\nu} h_{\mu\nu},
\end{eqnarray}
which is independent of the perturbed tetrad gauge freedom and, of course
invariant under $b_{\mu} \rightarrow b_{\mu} + \nabla_{\mu} \chi$; however
it is not under infinitesimal diffeomorphisms. On the other hand, it is
easy to verify that $\Psi^{\rm B}_{0}$ is invariant under ordinary gauge
transformations and infinitesimal diffeomorphisms (and independent of the
perturbed tetrad gauge freedom). In this manner, our quantities are, in
general, only invariant under both ordinary gauge transformations of the
vector potential perturbartions and rotations of the tetrad. \\[2em]

\begin{center}
{\uno APPENDIX B: EXACT ONE-FORMS ON THE PHASE SPACE}
\end{center}
\renewcommand{\theequation}{B\arabic{equation}}
\setcounter{equation}{0}
\vspace{1em}

The field variations appearing in the decoupled equation (3) correspond to
exact one-forms on the phase space.

The case of $\Psi^{\rm B}_{0}$ is trivial: $\Psi^{\rm B}_{0}$ is directly
the exterior derivative of the zero-form $\Psi_{0}$ (see last paragraph in
Sec. II).

The other cases are more elaborated. For example, ${\tilde \Psi}^{\rm
B}_{1}$ can be rewritten as:
\begin{equation}
    \tilde{\Psi}^{\rm B}_{1} \equiv \Psi^{\rm B}_{1} - \frac{3
        \Psi_{2}}{2 \phi_{1}} \phi^{\rm B}_{0} = \Psi^{\rm B}_{1} -
        \left( \frac{3\Psi_{2}}{2\phi_{1}} \phi_{0} \right)^{\rm B},
\end{equation}
since $\phi_{0} = 0$ in the background. Similarly, $\tilde{\kappa}^{\rm
B}$, and $\tilde{\sigma}^{\rm B}$ can be rewritten as:
\begin{eqnarray}
     \tilde{\kappa}^{\rm B} \!\! & = \!\! & \kappa^{\rm B} + \left[
     (D - 2 \epsilon - \rho) \frac{\phi_{0}}{2\phi_{1}} \right]^{\rm B},
     \nonumber\\
     \tilde{\sigma}^{\rm B} \!\! & = \!\! & \sigma^{\rm B} + \left[
     (\delta - 2 \beta - \tau) \frac{\phi_{0}}{2 \phi_{1}} \right]^{\rm B}.
\end{eqnarray}

Therefore, from Eqs.\ (B1) and (B2), $\tilde{\Psi}^{\rm B}_{1}$,
$\tilde{\kappa}^{\rm B}$, and $\tilde{\sigma}^{\rm B}$ correspond to the
exterior derivative of background zero-forms, and then correspond to exact
one-forms. A consequence of this property is that $(\Psi^{\rm B})$ is
made out of closed one-forms:
\begin{equation}
       (\Psi^{\rm B})^{\rm B} = 0.
\end{equation}

Similarly, the field variations  $\Psi^{\rm B}_{0}$, $\tilde{\Psi}^{\rm
B}_{1}$, $\tilde{\kappa}^{\rm B}$, and $\tilde{\sigma}^{\rm B}$ appearing
in Eq.\ (9) (and defined in Eqs.\ (39) of Ref. \cite{5}), correspond to
exact one-forms on the phase space.

The cases of $\Psi^{\rm B}_{0}$ and $\phi^{\rm B}_{0}$ in Eq.\ (8) are
also trivials: they are directly the variations of the corresponding
background quantity. \\[2em]


\begin{thebibliography}{}
\setlength{\itemsep}{-.50em}
\bibitem{1} C. Crncovi\'c and E. Witten, in {\it Three Hundred Years of
Gravitation}, edited by S. W. Hawking and W. Israel (Cambridge,
University Press. Cambridge, 1987).
\bibitem{2} J. Lee and R. M. Wald, J.\ Math.\ Phys.\ {\bf 31}, 3 (1990);
 G. A. Burnett and R. M. Wald, Proc.\ R.\ Soc.\ London {\bf
A430}, 57 (1990).
\bibitem{3} R. Cartas-Fuentevilla, Phys.\ Rev.\ D {\bf 57}, 3443 (1998).
\bibitem{4} G. F. Torres del Castillo and J. C. Flores-Urbina, Gen.\ Rel.\
Grav.\ {\bf 31}, 1315 (1999).
\bibitem{5} G. F. Torres del Castillo, J.\ Math.\ Phys.\ {\bf 5}, 649
(1988).
\bibitem{6} G. F. Torres del Castillo, J.\ Math.\ Phys.\ {\bf 37}, 4053
(1996).
\bibitem{7}  R. Cartas-Fuentevilla, J.\ Math.\ Phys.\, {\bf 41}, 7521
(2000).
\bibitem{8} G. F. Torres del Castillo, Gen.\ Rel.\ Grav.\ {\bf 22}, 1085
(1990).
\bibitem{9} R. M. Wald, Phys.\ Rev.\ Lett.\ {\bf 41}, 203 (1978).
\end{thebibliography}
\end{document}